\documentclass[%
 notitlepage,
 reprint,
 superscriptaddress,
 amsmath,amssymb,
 aps,
]{revtex4-2}

\usepackage{hyperref}
\usepackage{lipsum}
\usepackage{graphicx}
\usepackage{dcolumn}
\usepackage{bm}

\usepackage{epstopdf}
\usepackage{xcolor}

\newcommand{\bra}[1]{\langle #1 \vert}
\newcommand{\ket}[1]{\vert #1 \rangle}

\begin{document}

\preprint{APS/123-QED}

\title{Non-Markovian to Markovian decay in structured environments with correlated disorder}

\author{Mariana O. Monteiro}
\affiliation{%
 Instituto de F\'{i}sica, Universidade Federal de Alagoas, 57072-900 Macei\'{o}, AL, Brazil
}%

\author{\textcolor{black}{Nadja K. Bernardes}}
\affiliation{%
 Departamento de Física, Universidade Federal de Pernambuco, 50670-901 Recife, PE, Brazil
}%

\author{Eugene M. Broni}
\affiliation{%
 Instituto de F\'{i}sica, Universidade Federal de Alagoas, 57072-900 Macei\'{o}, AL, Brazil
}%

\author{Francisco A. B. F. de Moura}
\affiliation{%
 Instituto de F\'{i}sica, Universidade Federal de Alagoas, 57072-900 Macei\'{o}, AL, Brazil
}%

\author{Guilherme M. A. Almeida}
\affiliation{%
 Instituto de F\'{i}sica, Universidade Federal de Alagoas, 57072-900 Macei\'{o}, AL, Brazil
}%

\begin{abstract}
  Manipulating the dynamics of open quantum systems is a crucial requirement for large-scale quantum computers. Finding ways to overcome or extend decoherence times is a challenging task. Already at the level of a single two-level atom, its reduced dynamics with respect to a larger environment can be very complex. Structured environments, for instance, can lead to various regimes other than memoryless Markovian spontaneous emission. Here, we consider an atom
  coupled to an array of coupled cavities in the presence of on-site correlated disorder. The correlation is long-ranged and associated with the trace of a fractional Brownian motion following a power-law spectrum. With the cavity modes playing the role of the environment, we study the dynamics of the 
  spontaneous emission. We observe a change from non-Markovian to Markovian decay in the presence of disorder by tuning
  the correlation parameter. This is associated with a localization-delocalization transition involving the field modes. Two dissipative models that effectively reproduce the behavior of the non-Markovianity are discussed. 
  The dissipation dynamics of the atom can thus be used to extract information about the phase of the environment. Our results provide a direction in the engineering of disordered quantum systems to function as controllable reservoirs. 
\end{abstract}

\maketitle

\section{Introduction}


Understanding open quantum systems is pivotal
for the progress of quantum devices \cite{breuerbook}.
Quantum noise is generally viewed as a source of degradation of the quantum properties of the system, imposing significant challenges in achieving fault-tolerant quantum computation \cite{shor96,cai23}.
On the other hand, while every quantum system is open to some degree, it may not always be necessary to suppress the interaction between the system and its environment. Instead, one can leverage that interaction aiming at specific purposes \cite{koch16}. 
This perspective has been gaining traction in recent years, especially with the prospect of environment engineering to control dissipative quantum dynamics \cite{harrington22}. 
For instance, environment engineering allows the manipulation of the emission profile of a two-level atom \cite{myatt00,urrego18,longhi20,ferreira21}. 
Structured environments
are also a valuable resource for enhancing quantum technologies, allowing for the preparation of exotic quantum states \cite{mcendoo13,gonzalez17,lira24}, performing quantum simulation of open system dynamics \cite{barreiro11,semin20}, and much more \cite{harrington22,verstraete09}.
Many studies have addressed the transition between Markovian and non-Markovian regimes of dissipative dynamics in structured environments \cite{liu11,apollaro11,tufarelli14,francesco14,campbell18,gribben22,gaikwad24}. 
This kind of phenomenon has been recently observed on a superconducting qubit processor by tracking the evolution of an entangled two-qubit state, where one of the qubits interacted with a transmon playing the role of the environment \cite{gaikwad24}. 

A promising feature in the dynamics of open quantum systems comes from the perspective of quantum probing. 
Memory effects that may be present in the dissipative dynamics of a small system interacting with a much larger (and complex) environment, 
can be taken into account to learn  
about the latter using theoretical 
\cite{benedetti14,sone17,benedetti18,bina18,lyyra20,tamascelli20,blair24} and data-based methods \cite{luchnikov22,barr24}. 
These tasks aim at 
extracting relevant properties of the environment 
and its interaction with the main system
that otherwise would be difficult to measure directly (e.g., spectral densities).
Hence, 
involved many-body phenomena such as quantum and topological phase transitions are also 
encoded in dissipative quantum processes. 
\cite{haikka12,gessner14,lin16,giorgi19,yao20,rossini21}.


Most commonly,  non-Markovian dynamics are defined by quantum maps that cannot be divisible in other quantum channels \cite{rivas14}. However, dynamics that do not satisfy the semigroup property and, consequently, their master equations do not have a representation in Lindblad form are also called non-Markovian \cite{alickibook}.

One key signature of non-Markovianity is the occurrence of information backflow into the main system 
\cite{madsen11,pastawski11,man15,fang18,shen19,li24}. 
This is ultimately determined by the profile of the spectral density of the environment. 
Structured environments, such as 
photonic crystals \cite{lambropoulos00}, exhibit a rich variety of spectral functions, leading to diverse non-Markovian dynamical regimes.
As such, strong memory effects in a dissipative two-level atom can be induced by either controlling the formation of bound states \cite{longhi07,francesco14} via the coupling strength or by 
inducing the presence of localized states in the environment \cite{sapienza10, lorenzo17, lorenzo17-2, cosco18}. In an amplitude damping channel \cite{lorenzo17}, deviations from the standard exponential decay (Markovian regime) are expected in the presence of disorder. 
%

%
In this work, we consider a coupled-cavity array (CCA) containing a two-level atom trapped in the middle cavity undergoing Jaynes-Cummings interaction, as depicted in Fig \ref{fig1}(a).
Coupled-cavity systems have attracted a great deal of interest for their promising applications in simulating quantum phase transitions \cite{hartmann06,greentree06}, realizing quantum communication protocols \cite{almeida16, mendonca20}, and more \cite{meher22}. In the so-called Jaynes-Cummings-Hubbard model \cite{greentree06}, each cavity contains a two-level atom. Here, instead, all the cavities are empty, except one, 
such that the environment is structured according to the free-field normal modes.   
We consider the frequencies through the CCA to follow a disordered series embedded with long-range correlations obeying a power-law spectrum of the form $k^{-\alpha}$, $\alpha$ being the correlation degree \cite{moura98}. This model is known to support a localization-delocalization transition at $\alpha=2$, which, as we demonstrate, decreases the degree of the non-Markovianity of the decay dynamics despite the existence of localized modes. 
In summary, we show that Markovian (non-Markovian) decay is associated with strong (weak) long-range correlations, with the global disorder strength held constant. 
Hence, on the one hand, 
we can control the non-Markovianity in the emission dynamics by tuning the disorder correlation exponent $\alpha$. On the other hand, the evolution of the atom (a qubit) reflects the localization-delocalization phase transition occurring in the environment.

\section{Methods
}
\subsection{Hamiltonian model}

\begin{figure}[t!]
    \includegraphics[width=0.45\textwidth]{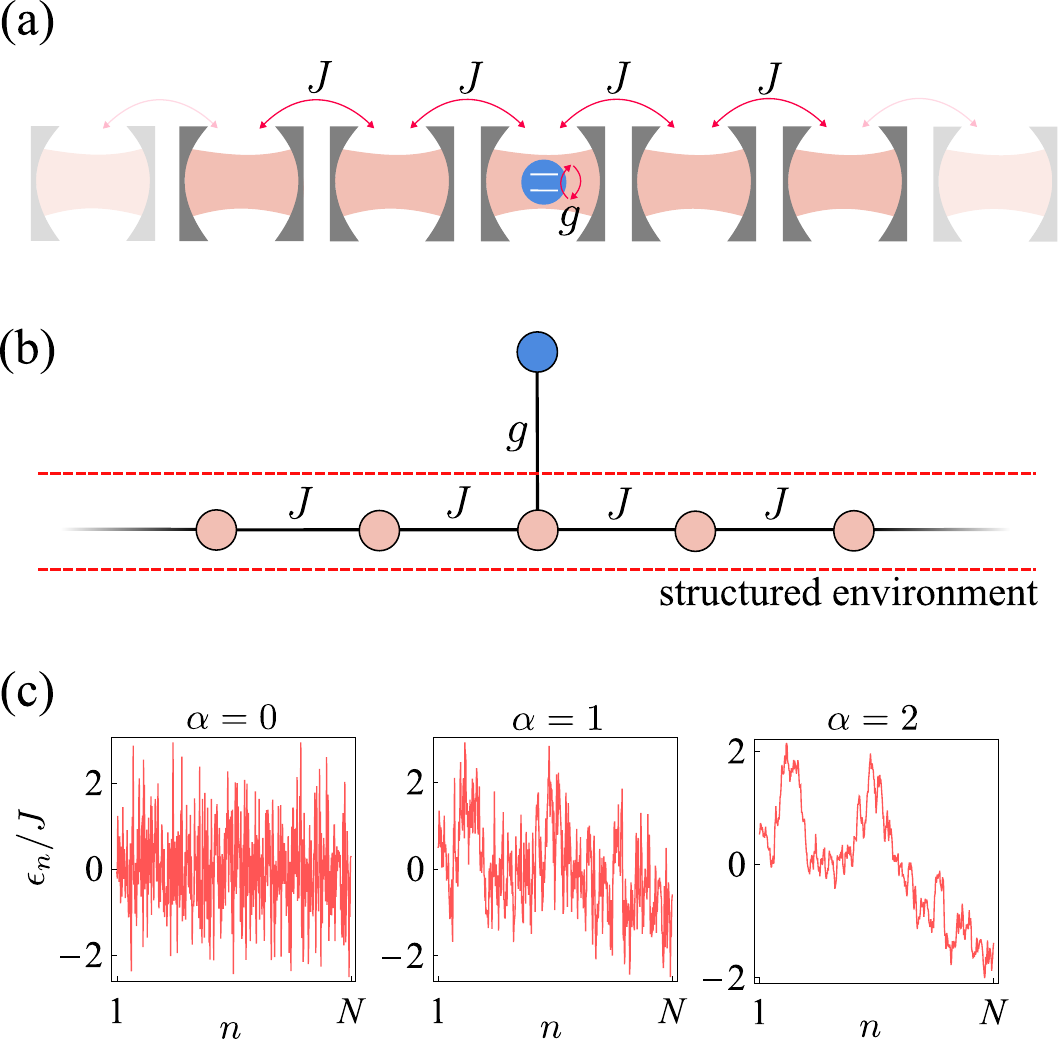}
        \caption{(a) Array of optical cavities coupled with tunneling rate $J$. A two-level atom is confined at the center 
        and couples to the field mode at rate $g$. (b) Equivalent graph in the single-excitation subspace. In this work, we 
     follow the dynamics of the atomic spontaneous emission into the cavity array, initially in the vacuum, which serves as an environment structured with correlated disorder in the cavity frequencies $\epsilon_n$. (c) Sample of the correlated series
        generated by Eq. (\ref{epsilon}) for
        different $\alpha$, which controls the degree of long-range correlations.}
        \label{fig1}    
    \end{figure}

Consider a CCA
described by the Hamiltonian $\hat{H}=\hat{H}_0+\hat{H}_I$, where ($\hbar = 1$)
\begin{equation}\label{H0}
    \hat{H}_{0} = \sum^{N-1} _ {n=1}\left[\epsilon_{n}\hat{a}^{\dagger} _ {n}\hat{a} _ {n} + 
     J(\hat{a} _ {n}\hat{a}^{\dagger} _ {n+1}+\hat{a}^{\dagger} _ {n}\hat{a} _{n+1}) \right]
\end{equation}
represents photon tunneling through
the $N$ cavities.
The annihilation (creation) operator $\hat{a}_{n}$ ($\hat{a}^{\dagger}_{n}$) acts in the $n$th cavity and each one supports a mode with frequency $\epsilon_n$. A single two-level system is confined in the central cavity (we take $N$ odd), which we label as $c\equiv (N+1)/2$. The full scheme is depicted in Fig. \ref{fig1}(a). The emitter interacts with the field mode via the Jaynes-Cummings Hamiltonian (in the rotating wave approximation) 
\begin{equation}
    \hat{H}_I = w_{a} \hat{\sigma}_{+}\hat{\sigma}_{-} + g (\hat{\sigma}_{+}\hat{a}_{c} + 
    \hat{\sigma}_{-}\hat{a}^{\dagger}_c),
\end{equation}
where $g$ is the coupling rate, $w_a$ the atomic frequency and 
$\sigma_{+}$ ($\sigma_{-}$) the atomic raising (lowering) operator defined as $\sigma_{+} = \ket{e}\bra{g}$. 
Our goal here is to track the dynamics of the initial state
$\ket{e}\ket{\mathrm{vac}}$, namely
the atom in the excited state with all the cavity modes in the vacuum.
As the full Hamiltonian commutes with the total number of excitations, it suffices to work within the single excitation subspace [see Fig. \ref{fig1}(b)] spanned by the set $\lbrace \ket{e}\ket{\mathrm{vac}},\lbrace \ket{g}\ket{n} \rbrace \rbrace$, with  
$\ket{n}$ representing a single photon at cavity $n$. 

Another way to cast the full Hamiltonian is in terms of the normal modes of the free CCA, i.e., the environment, satisfying $\hat{H}_0\ket{\phi_k}=\omega_k\ket{\phi_k}$, with $\ket{\phi_k} = \sum_n v_{k,n}\ket{n}$. The Hamiltonian thus reads 
\begin{equation}\label{Hnormal}
\hat{H} = w_{a} \hat{\sigma}_{+}\hat{\sigma}_{-} + \sum_{k}\omega_k\hat{\phi_k}^{\dagger}\hat{\phi_k}+\sum_k g_k (\hat{\phi_k}\hat{\sigma}_{+}+\hat{\phi_k}^{\dagger}\hat{\sigma}_{-}),
\end{equation}
where $\hat{\phi_k}^{\dagger} = \ket{\phi_k}\bra{\mathrm{vac}}$ and $g_k = gv_{k,c}=g\langle c \vert \phi_k \rangle$ denoting the effective coupling between the atom and the environment mode $k$. We assume $g_k$ is real for simplicity.

The scheme outlined in Fig. \ref{fig1}(b) was studied both in the absence \cite{longhi07,francesco14} and in the presence \cite{lorenzo17} of disorder in $\epsilon_n$, in the weak-coupling regime $g\ll J$. In \cite{lorenzo17}, static disorder described by random cavity detunings was shown
to induce quantum non-Markovianity with increasing disorder strength. It is widely known that uncorrelated disorder in 1D entails Anderson localization of all the modes. This results in a non-homogeneous distribution of the couplings $g_k$. For strong disorder, 
the atom can engage in a Jaynes-Cummings dynamics with the mode 
having the highest amplitude in cavity $c$.
In contrast, a homogeneous CCA (or any tight-binding chain) provides a flat spectral density in the bulk, which can be envisaged from its cosine dispersion law and delocalized Bloch modes. Thus, standard (Markovian) emission takes place with the atomic population decaying as  $p_e(t) = e^{-g^2t / J}$ for $g\ll J$ \cite{francesco14}.

    The two scenarios described above, namely undamped Jaynes-Cummings dynamics (vacuum Rabi oscillations) and pure exponential decay, embody our context's limiting non-Markovian and Markovian regimes. Anything between will be quantified using a proper non-Markovianity metric, which will be introduced shortly. The general question we want to address now is whether roaming between those two regimes under a fixed global disorder strength is possible. 

\subsection{Long-range correlated disorder}

    Anderson localization in 1D systems can be effectively weakened by introducing correlations in the disorder \cite{izrailev12}.  
    In this work, we assume that the local potential follows a 
    random series carrying long-range correlations. We can generate it according to the trace of a fractional Brownian motion with power-law spectrum $\propto k^{-\alpha}$ \cite{moura98}:
    \begin{equation}
         \epsilon_{n} = \sum_{k=1}^{(N+1)/2} k^{-\alpha/2}\cos\left ( \frac{2\pi nk}{L} + \varphi_{k} \right ),
         \label{epsilon}
    \end{equation}
where $k=1/\lambda$, $\lambda$ being the wavelength of the modulation profile, and ${\varphi_{k}}$ are random phases falling between $[0,2\pi)$. 
Hereafter we will fix the global disorder strength by normalizing $\epsilon_n$ such that $\langle \epsilon_n \rangle=0$ and $\mathrm{var}(\epsilon_n)=1$ for each realization of the series. 
The exponent $\alpha$ establishes the degree of correlation of the disorder, controlling the series trend, as shown in Fig. \ref{fig1}(c).
Note that white noise resembling uncorrelated disorder is recovered for $\alpha=0$. This model predicts a localization-delocalization transition at $\alpha=2$. The energy landscape acquires a self-similar and persistent character for $\alpha>2$, which is related to the vanishing of the Lyapunov
coefficient around the center of the band.
In the following section, we will see how it affects the non-Markovian character of the spontaneous emission of the atom.

\subsection{Non-Markovianity quantifier}

A large body of literature is dedicated to measures and witnesses of non-Markovianity. We refer to the reader the reviews in Refs. 
\cite{rivas14, breuer16, vega17}. To decide which one to pick, 
let us ponder what kind of dissipation channel we have.
The state ket at an arbitrary time $t$ reads 
\begin{equation}\label{psievo}
\ket{\psi (t)} = \hat{U}(t)\ket{\psi(0)}= f_e(t)\ket{e}\ket{\mathrm{vac}}+ \sum_k f_k(t)\ket{g}\ket{\phi_k},
\end{equation}
with $ \hat{U}(t) = e^{-i \hat{H}t}$ being the unitary time evolution operator and $f_e(t), f_k(t)$ the atomic and field amplitudes, respectively.
For the initial state $\ket{\psi(0)}=\ket{e}\ket{\mathrm{vac}}$, the
reduced atomic density matrix becomes $\rho_A(t) = \mathrm{Tr}_E \lbrace \ket{\psi(t)}\bra{\psi(t)} \rbrace = \mathrm{diag}(p_e(t),1-p_e(t))$,
where $E$ stands for the CCA environment (field degrees of freedom) and $p_e(t) \equiv |f_e(t)|^2$.
Therefore, the system goes through
an amplitude damping channel. 

A handy non-Markovian witness in this case is associated to the change of the volume
of physical states that are dynamically accessible to a system \cite{lorenzo13}. It is possible to show that this quantity, in the case of the amplitude damping channel, is associated to any increase in $p_e(t)$ over time.

Formally, a proper measure can be established by tracking all the positive slopes of $\partial_t p_e(t)$ over time. 
Here, we will rely on the same formula used in Ref. \cite{lorenzo17}, defined as $\mathcal{N} = \mathcal{N}_V/|\widetilde{\mathcal{N}}_V|$, where
\begin{equation}\label{NV}
\mathcal{N}_V = \int_{\partial_t p_e(t)>0}dt  \frac{dp_e^2(t)}{dt}.
\end{equation}
%
The denominator $\widetilde{\mathcal{N}}_V$ is defined
similarly to Eq. (\ref{NV}), but with the integral evaluated over
$\partial_t p_e(t)<0$.
This is included to prevent the divergence of $\mathcal{N}_V$ associated with undamped Rabi oscillations. Consequently, $\mathcal{N}$ goes from $\mathcal{N}=0$ (Markovian decay), when $\widetilde{\mathcal{N}}_V$ diverges, to $\mathcal{N}=1$ (non-Markovian decay).

The measure $\mathcal{N}_V$ is
inspired by the geometrical approach to non-Markovianity introduced in \cite{lorenzo13}, rooted on the volume of states (here denoted by $p_e^2$) that can be accessed during evolution. 
The formula can be simplified considering that $p_e(\infty)=0$ for an amplitude damping channel. Then, it can be shown that $\widetilde{\mathcal{N}}_V = \mathcal{N}_V+1$, such that $\mathcal{N}=\mathcal{N}_V/(\mathcal{N}_V+1)$ \cite{lorenzo17}. We can simplify it further with the aid of the fundamental theorem of calculus, which leads to
$\mathcal{N}_V = \sum_M p_e^2(t_M) - \sum_m p_e^2(t_m)$, where $t_M$ and $t_m$ are the times corresponding to the maxima and minima of $p_e^2(t)$, respectively.


\section{Results and discussion}

We now investigate the dynamics of the atom interacting with the structured CCA environment (initially in the vacuum state). We will see that the influence of the correlated disorder generally benefits memoryless, Markovian dynamics given the frequency of the atom $\omega_a$ is tuned to the center of the band. 
  Following that, we will quantify the associated non-Markovianity using $\mathcal{N}$ and discuss a couple of effective models that phenomenologically capture its behavior.

\subsection{Spontaneous emission and non-Markovianity}

 \begin{figure}[t!]
    \includegraphics[width=0.45\textwidth]{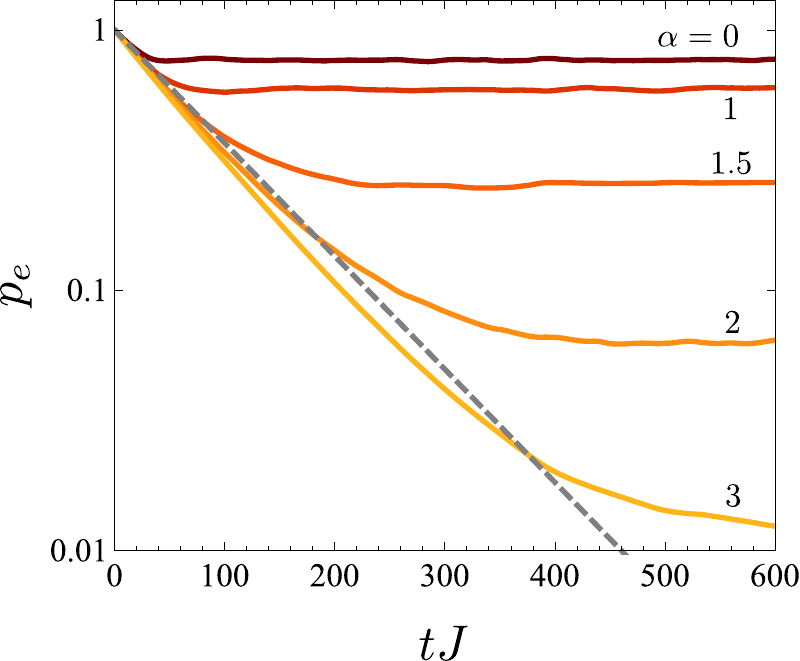}
        \caption{Time evolution of the atomic excitation probability $p_e$ (in log-lin scale) for some values of $\alpha$. System parameters are $g=0.1J$, $w_a=0$, $N=6201$, and each curve represents the average outcome of $10^3$ independent realizations of the disorder. For comparison, the dashed curve represents
        $p_e(t) = e^{-g^2t / J}$, which is the pure Markovian decay that the atom would experience in the case of a flat spectral density supported by a homogeneous CCA.}
        \label{fig2}    
    \end{figure}

Considering the evolved state expressed in Eq. (\ref{psievo}) the time-dependent Schrödinger equation leads to:
\begin{align} \label{sch}
\frac{df_e(t)}{dt} &=-i\omega_a f_e(t) -i\sum_kg_k f_k(t),\\
\frac{df_k(t)}{dt} &=-i\omega_k f_k(t) -i g_k f_a(t).
\end{align}
Solving for $f_k(t)$ under $f_k(0)=0$ and replacing into Eq. (\ref{sch}) we obtain
\begin{equation} \label{fedot}
\frac{df_e(t)}{dt} = -i\omega_a f_e(t)-\int_{0}^{t}\sum_k g_k^2 e^{-i\omega_k (t-t')}f_e(t')dt' ,
\end{equation}
which is the starting point for problems involving spontaneous emission \cite{lorenzo17, mouloudakis22-2}. From here, 
if one is interested in studying the emitter coupled
to a continuum, then $\sum_k g_k^2 \rightarrow \int G(\omega)d\omega$, where $G(\omega)\equiv g^2(\omega)\rho(\omega)$ is the spectral density and $\rho(\omega)$ is the density of states.
Once $G(\omega)$ is specified for the problem at hand, the solution for $f_e(t)$ can be sought via the Laplace transform (see, e.g., Refs. \cite{lambropoulos00,mouloudakis22-1, mouloudakis22-2}).

The spectral density is a key quantity in the study of open quantum systems
as it contains all the relevant information about the interaction between the system and its environment. Indeed, the shape of $G(\omega)$ determines what kind of dissipative dynamics the atom will undergo. A flat spectral density generates
Markovian dynamics, characterized by memoryless exponential decay of the atomic excitation. 
In practice, this regime typically holds when $G(\omega)$ varies slowly over a frequency range and the system-environment coupling is weak such that $G(\omega)$ can be effectively regarded as constant.
If $G(\omega)$ has a peak around some $\omega\approx \omega_a$, then memory effects emerge in the form of damped Rabi oscillations between the atom and the environment. 
For an atom inside a high-$Q$ cavity, for example, such a peak 
has a narrow lineshape. Yet, the wings of the distribution may extend to infinity, rendering the full decay of the atom at some point. A common approach in this case is to assume a Lorentzian spectral density \cite{lambropoulos00}.
%
%
The existence of discontinuities in $G(\omega)$, as found in photonic band-gap materials \cite{sajeev90,kofman94,lambropoulos00}), results in more involved dynamics, with the contribution of bound states. In these cases, persistent quantum oscillations preclude the system from reaching equilibrium with its environment \cite{xiong15}.    

In our system, uncorrelated disorder ($\alpha=0$) makes 
every field mode exponentially localized at a given location. Following the reasoning above and inspecting Eq. (\ref{Hnormal}), a very strong disorder yields a situation similar to that of a high-$Q$ cavity, provided $\omega_a$ is resonant to the mode $\omega_k$ having the highest overlap with the central cavity $c$ \cite{lorenzo17}. 

Because the parameters of our system fluctuates
and we are mostly interested in the \emph{average} behavior of $p_e$ and $\mathcal{N}$ over many independent disorder realizations (typically $10^3$) we will 
solve the time evolution of the system 
using a high-order Taylor expansion of the evolution operator:
\begin{equation}
\hat{U}(\Delta t) = \exp(-i\hat{H}\Delta t) = 1 + \sum_{l=1}^{n_0} \frac{(-i\hat{H}\Delta t)^l}{l!}.
\end{equation}
thus, the state ket can be obtained at recursively any time $t$. We use a time step of \(\Delta t = 0.1J\) and truncate the sum at \( n_0 = 12 \), which is enough to preserve
the norm $\sim 1-10^{-8}$ during the dynamics. This method allows us to account for large CCAs, thereby making the simulation free from boundary effects.
Other quantities based on the spectral structure of free-field Hamiltonian $\hat{H}_0$ (environment) will be obtained via its exact numerical diagonalization.

  \begin{figure}[t!]
    \includegraphics[width=0.45\textwidth]{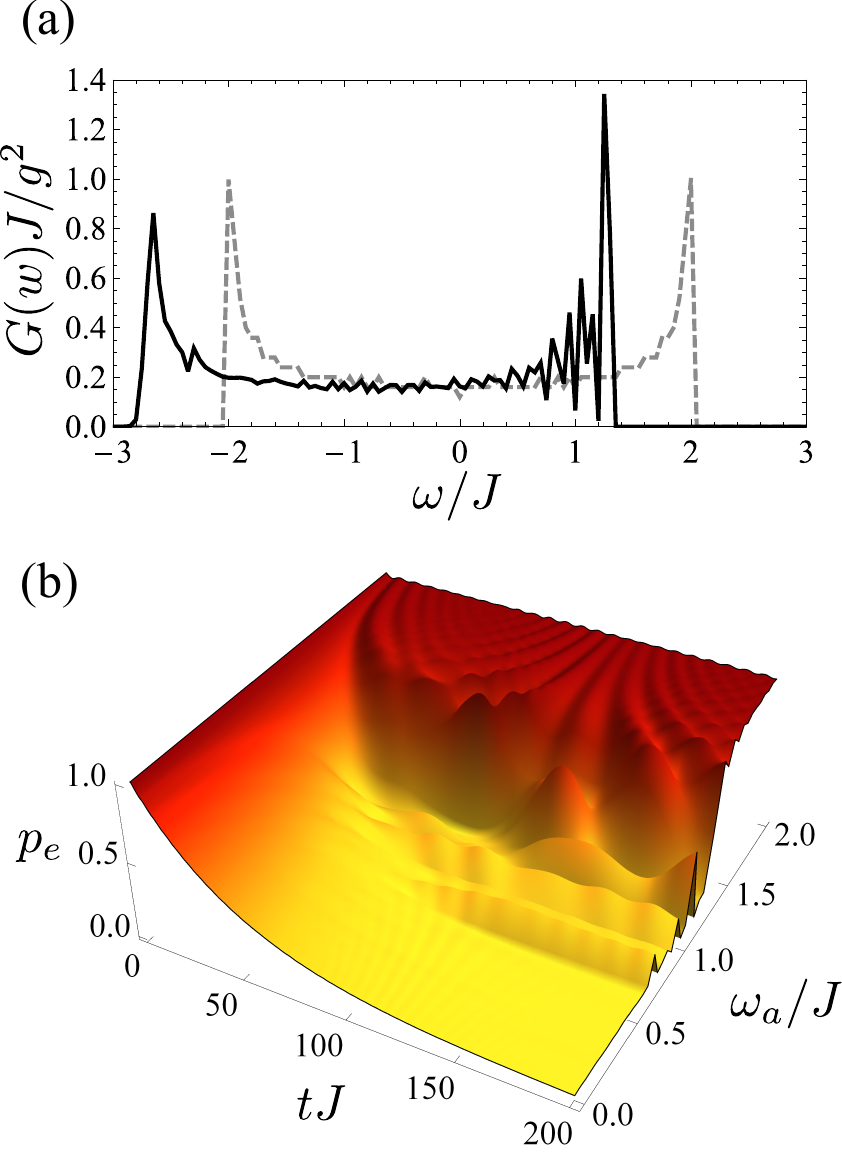}
        \caption{(a) Typical realization of the spectral density $G(\omega)=|g(w)|^2\rho(\omega)$ versus $\omega$ for $\alpha = 3$ (solid line) and the homogeneous CCA (dashed line) for comparison, obtained via exact numerical diagonalization of $H_0$ considering $N=1001$. \textcolor{black}{$G(w)$ is calculated by summing all $g_k^2$
        corresponding to 
        the window $[\omega_k-0.05J,\omega_k+0.05J]$ and dividing by the bin size 
        $0.1J$}.
        (b) Corresponding time evolution of the atomic population $p_e$ against $\omega_a$ considering $g=0.1J$. When the atomic frequency $\omega_a$ is tuned to the center of the band, the emission is Markovian (since $G(\omega) \approx \mathrm{constant}$ in the weak coupling regime $g \ll J$). Localized field modes populate higher frequencies, inducing memory effects. The atomic excitation freezes when $\omega_a$ is set beyond the band edge.}
        \label{fig3}    
    \end{figure}

     \begin{figure}[t!]
    \includegraphics[width=0.45\textwidth]{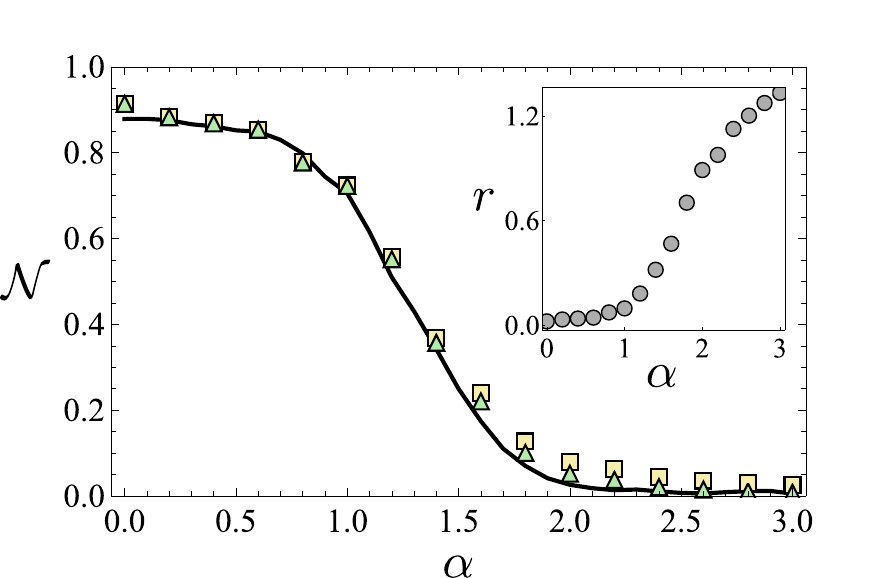}
        \caption{Non-Markovianity $\mathcal{N}$ versus disorder correlation parameter $\alpha$ for $\omega_a = 0$. Solid line depicts the averaged quantity obtained from the time evolution of the full Hamiltonian with $N=6201$ and $g=0.1J$ over an ensemble of $10^3$ realizations of disorder.  Symbols are fittings originated from the effective models, 
        namely (squares) the emitter in contact with a Markovian bath plus 
        an auxiliary mode; and (triangles) the emitter interacting with a Lorentzian bath (cf. Fig. \ref{fig5}).
        Inset: Averaged $r = \gamma /g_\ell$, which feeds the effective models, evaluated for $10^3$ realizations of the free-field Hamiltonian $\hat{H}_0$, with $N=1001$.
        The decay rate $\gamma = \gamma(\omega_a = 0)$
was obtained 
by summing all $g_k^2$
        corresponding to 
        the window $[\omega_a-0.05J,\omega_a+0.05J]$ and dividing by the bin size 
        $0.1J$. The effective coupling was set to $g_\ell = g/\sqrt{\xi}$, 
        with unit proportionality constant for both models. 
}
        \label{fig4}    
    \end{figure}

The dynamics of the atomic population $p_e$ is displayed in Fig. \ref{fig2} for a few values of the correlation parameter $\alpha$. 
We readily see that the case in which the disorder is uncorrelated ($\alpha=0$) is strongly characterized by a population trapping. Increasing $\alpha$ gradually leads to a nearly exponential decay of the excitation. 
The curves are plotted in log-linear scale for better visualization.
The dashed straight line in the figure accounts 
for the fully Markovian behavior given by $p_e(t) = e^{-g^2t / J}$ (see e.g. \cite{francesco14}), obtained for a continuum of modes of an infinite and homogeneous CCA ensuing a bath with flat spectral density. Henceforth, we are using $g=0.1J$ (weak coupling regime). The time window $tJ\in[0,600]$ used in 
the simulations, besides securing no reflections from the boundaries,
it is adequate to obtain the saturated behavior of the 
ensemble averages of $p_e$. 
%
Note that for $\alpha \geq 2$ there is still some contribution 
reminiscent of bound states to the dynamics, as indicated 
by the deviation from the exponential at longer times. 
Yet, it is remarkable that in the decay process more than $90\%$ of the atomic excitation is released 
in close resemblance to the Markovian benchmark despite the fixed disorder strength set by $\mathrm{var}(\epsilon_n)=1$.   

To understand how the CCA environment behaves for large values of $\alpha$ it is useful to analyze the spectral properties and the dynamics for single realization of the disorder. 
In Fig. \ref{fig3}(a) the spectral density $G(\omega)$ is shown for a typical sample when $\alpha=3$. The plot reveals a clear similarity 
to the spectral density of the
 homogeneous CCA (dashed line), featuring the van-Hove singularities near the band edges. 
 However, two key differences can be identified.
  First, $G(\omega)$ becomes less flat as we move away from the center of the band, suggesting the presence of localized states. Second, the center is shifted from $\omega = 0$. The disorder sample was intentionally selected to highlight this. In fact,
the band is always offset to some degree (either to the left or to the right) for distinct realizations of the disorder. 
Hence, if the frequency of the emitter is fixed at $\omega_a=0$ there is a chance that localized 
modes are present in nearby frequencies. This explains the averaged behavior of $p_e$ displayed in Fig. \ref{fig2} for large values of $\alpha$. For all practical purposes, however, the decay dynamics can be 
classified as Markovian as we will see shortly. 

It is also interesting to look what happens when $\omega_a$ assumes other values across the band.
Figure \ref{fig3}(b) shows the time evolution of $p_e$ for a wide range of values of $\omega_a$ for the same disordered sample. We see that atomic decay is of exponential form up to $\omega_a \approx 0.6J$. 
For higher frequencies, memory effects (information backflow) is apparent, until the atomic excitation becomes essentially trapped as soon as $\omega_a$ crosses the band edge, as expected. 

As our goal is to observe the transition between the non-Markovian dynamics associated to the dominant presence localized states for $\alpha=0$ toward the Markovian regime, we will keep $\omega_a=0$ throughout. This transition is properly accounted for via the non-Markovianity measure introduced earlier: $\mathcal{N} = \mathcal{N}_V/|\widetilde{\mathcal{N}}_V|$ [see Eq. (\ref{NV})]. The results are displayed in Fig. \ref{fig4} (solid line), for the same parameters as in Fig. \ref{fig2}, with the $\mathcal{N}$ being numerically evaluated for each realization of the disorder and averaged afterwards. Therein, we note that the onset of the transition towards the Markovian regime occurs between $\alpha=1$ and $\alpha=2$. Studies have suggested that within this range of values the participation ratio behaves sublinearly with $N$ \cite{santos06,mendes19}.
Therefore, the localization length becomes larger although the modes are not truly delocalized yet. This is guaranteed only for $\alpha>2$ which marks the transition. The delocalized states populate a frequency range around the center of the band and are surrounded by mobility edges as shown in \cite{moura98}. As in \cite{lorenzo17}, where the same measure of non-Markovianity was applied to a disordered CCA with no correlations, here we do not have a threshold separating both regimes.
Here, the quantity $\mathcal{N}$ is always finite also due to the fact that we are averaging it over an ensemble. In fact $\mathcal{N}=0$ only in the case of an infinite homogeneous CCA. 
Notwithstanding, we can reasonably establish that the Markovian regime is reached for $\alpha>2$, as supported by the interaction of the atom with delocalized modes embodying a flat spectral density. 

\subsection{Effective models}

 \begin{figure}[t!]
    \includegraphics[width=0.45\textwidth]{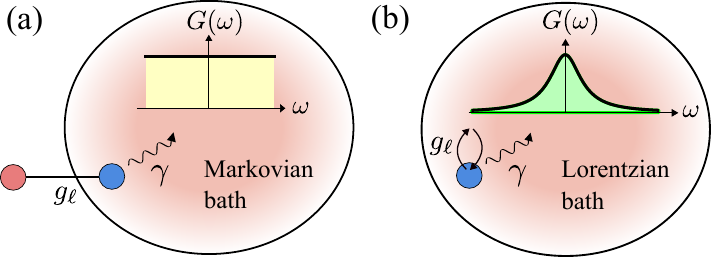}
        \caption{Effective schemes that capture the non-Markovianity behavior displayed in Fig. \ref{fig4}. In (a) the emitter is under the influence of a Markovian (flat) bath decaying at rate $\gamma$ while also coupled at $g_\ell$ to an auxiliary mode (a free-field mode) with local frequency $\omega_\ell$ and not in direct contact with the bath.
In (b) the system interacts with a Lorentzian bath whose shape parameters read from both $\gamma$ and $g_{\ell}$. 
}
        \label{fig5}    
    \end{figure}

We now introduce two phenomenological models able to
capture the dependence of the non-Markovianity $\mathcal{N}$ on $\alpha$ (Fig. \ref{fig4}). Both models are illustrated in \ref{fig5}. We stress that we do not want to reproduce the exact dynamics of $p_e$. Rather, we need something that accounts for the competition between undamped Rabi oscillations and exponential decay having the localization degree of the modes that compose the CCA environment as input.


First, as done in Ref. \cite{lorenzo17}, let us imagine an emitter in contact with a Markovian bath, decaying at rate $\gamma$ while at the same time interacts with an external mode $\ell$ via the coupling strength $g_\ell = g\langle c \vert \psi_\ell \rangle$ [see Fig. \ref{fig5}(a)]. This mode is not in \emph{direct} contact with the bath. Intuitively, we realize that the ratio $r= \gamma / g_\ell$ will determine how fast the excitation roams between the two states before being fully released into the environment. As far as non-Markovianity is concerned, we can relate that to what happens in our system as the disorder becomes more correlated. 
%

Such a dissipative two-level model can be worked out from the Hamiltonian in Eq. (\ref{Hnormal}) by assuming that the atom is strongly coupled to a specific mode $\ell$ via $g_\ell$, with the remaining ones forming the effective Markovian bath.
A master equation in Lindblad form for the reduced state $\rho_{e,\ell}$ of the emitter and mode $\ell$
can thus be derived as (see \cite{lorenzo17} for details)
\begin{equation}\label{master}
\frac{d \rho_{e,\ell}}{dt} = -i[\hat{H}_\ell, \rho_{e,\ell}(t)] + \gamma\left( \hat{L} \rho_{e,\ell} \hat{L}^\dagger - \frac{1}{2} \left\{ \hat{L}^\dagger \hat{L}, \rho_{e,\ell} \right\}\right),
\end{equation}
where $\hat{H}_\ell = \omega_a \hat{\sigma}_+\hat{\sigma}_- + \omega_\ell \hat{\phi}_{\ell}^\dagger \hat{\phi}_{\ell} + g_\ell (\hat{\sigma}_+ \hat{\phi}_{\ell} + \mathrm{h.c.})$ 
and the jump operator $\hat{L}= \hat{\sigma}_-$. 
Here, the dissipation rate is defined as $\gamma = \gamma(\omega_a) = \pi \sum_{k\neq \ell} g_k^2\delta(\omega_k-\omega_a)$, which basically depends on the spectral density evaluated at $\omega = \omega_a$ excluding the contribution from mode $\ell$. 

The atomic probability that results from the master equation above reads (we set $\omega_a = \omega_\ell$ for simplicity)
\begin{equation}\label{model1}
p_e(t) =e^{-\frac{\gamma}{2} t}\left[ \cos\left(\frac{\Delta \gamma}{4} t\right)- \frac{1}{\Delta} \sin\left(\frac{\Delta \gamma}{4}  t \right) \right]^2 ,
\end{equation}
where $\Delta = \sqrt{16-r^2}$. The expression above is valid for $r\neq 4$. When $r=4$, we have $p_e(t) = e^{-\frac{\gamma}{2} t}(1-\gamma t / 4)^2$.

Let us now discuss a second phenomenological model that 
does not require coupling to an external mode. As such, the information backflow dynamics (here occurring in the form of Rabi oscillations) must be encoded in the spectral density of bath. This leads us to think of a scenario similar to that of an atom inside a high-$Q$ resonator, whose spectral density is usually given by a Lorentzian peaked at $\omega_0$:
\begin{equation}
G(\omega) = \frac{g_\ell^2}{\pi}\frac{\frac{\gamma}{2}}{(\omega-\omega_0)^2+(\frac{\gamma}{2})^2}
\end{equation}
where the dissipation rate $\gamma$ enters as the width of the distribution. Note that the Markovian limit can be reached 
 by making $\gamma \rightarrow \infty$ ($r\rightarrow \infty$). 
For the atomic frequency in resonance with the peak center, $\omega_a = \omega_0$, an analytical solution for $p_e$ 
can be obtained \cite{lambropoulos00}:
\begin{equation}\label{model2}
p_e(t) = \frac{1}{2}e^{-\frac{\gamma}{2} t}\left[1+\cos\left(\frac{\Delta \gamma}{2r}t\right)\right].
\end{equation}

With both expressions for $p_e$ at hand and recalling that 
$\mathcal{N}_V = \sum_M p_e^2(t_M) - \sum_m p_e^2(t_m)$, 
our goal here resumes to find all the maxima, because $p_e(t_m)=0$ for all $t_m$
in Eqs. (\ref{model1}) and (\ref{model2}). After some calculations, one finds 
$
\mathcal{N}_V = e^{-rt_0}(e^{4\pi r / \Delta}-1)^{-1},
$ \cite{lorenzo17}
for the first model [see Fig. \ref{fig5}(a)], with $t_0 = 4\Delta^{-1}\arctan{\lbrace 2\Delta r/(r^2-\Delta^2)\rbrace}$. This expression for $\mathcal{N}_V$ is valid for $r\in [0,2\sqrt{2})$. 
%
%
Considering the second model [Fig. \ref{fig5}(b)], from Eq. (\ref{model2}) we get $\mathcal{N}_V =(e^{4\pi r / \Delta}-1)^{-1}$, valid for $r\in [0,2)$. 

The connection between the effective models and the actual CCA system is made by first defining the mode $\ell$ of $H_0$ [Eq. (\ref{H0})] as the one 
with the largest factor $|g_k|/|\omega_k-\omega_a|$.
The argument is that the localized nature of the free-field modes favors the pairing between $\ket{e}\ket{\mathrm{vac}}$ and a given $\ket{g}\ket{\phi_k}$ offset by the detuning between them. 
From here, we can define $g_\ell\propto g/\sqrt{\xi}$, where $\xi=(\sum_n |v_{\ell,n}|^4)^{-1}$ is the participation ratio, grounded on the analogy between cavity volume and localization length \cite{lorenzo17}. The participation ratio expressed in this manner ranges from $1$ to $N$, for fully localized and delocalized states, respectively. 

Further, we numerically evaluate the dissipation rate $\gamma = \gamma (\omega_a)$ at the center of the band (that is $\omega_a=0$) to ensure the influence of delocalized states as $\alpha$ increases.
Despite Eqs. (\ref{model1}) and (\ref{model2}) being valid for perfect resonant
conditions in the effective models, we still have $\omega_a \approx \omega_\ell$ on average. Most of all, the non-Markovianity is not dramatically affected by a small detuning.

Once $r(\alpha)=\gamma (\alpha)/ g_{\ell} (\alpha)$ is obtained by averaging over a large ensemble of realizations of the disorder (see inset of Fig. \ref{fig4}), we can feed it into the simplified 
formula for the non-Markovianity, $\mathcal{N} = \mathcal{N}_V/(\mathcal{N}_V+1)$. The results are shown in Fig. \ref{fig4}, where each of the two sets of symbols represents a given phenomenological model. Both show remarkable agreement with the original data (solid line), especially the model featuring the Lorentzian bath [Fig. \ref{fig5}(a)] for higher values of $\alpha$. This happens due to absence of the factor $e^{-rt_0}$ in the corresponding expression for $\mathcal{N}_V$.

Finally, note that for $\mathcal{N}$
evaluated using both effective models, it is implied that $p_e(\infty)=0$. Indeed, the corresponding spectral densities have wings that extend to infinity. Yet, we were able to reproduce the behavior of the non-Markovianity of the CCA evolving up to a time $tJ= 600$. 
This can be explained with regard to saturated behavior of the averaged 
$p_e$ seen in Fig. \ref{fig2}. Even though during such a time frame the excitation is not fully released (and never will given the discreteness of the spectrum of $H_0$), the average $r=\gamma / g_\ell$ contains the necessary information to predict the kind of dynamics that the atom will undergo. 


\section{Conclusions and outlook}

In summary, we have investigated the dynamics of the spontaneous emission of a two-level atom into a disordered environment featuring correlated disorder, represented by the CCA modes. By quantifying the non-Markovianity using $\mathcal{N}$, we showed that increasing the correlation parameter $\alpha$ leads to a nearly exponential decay. The transition significantly takes place between $\alpha=1$ and $\alpha=2$, after which we establish the existence of a Markovian regime in association with the flat spectral density in the center of the band induced by the presence of delocalized states \cite{moura98}. As shown in previous papers \cite{lorenzo17}, we confirm that a disordered phase and, consequently, delocalization leads to Markovian dynamics.

In addition, two simple phenomenological models were discussed: a two-level system in contact with a Markovian bath and an emitter interacting with a Lorentzian spectral density. Using the localization properties of the CCA as inputs, these models could reproduce the non-Markovianity behavior quite accurately through analytical formulas.
Therefore, in general, such a framework provides a powerful tool to explore a variety of spontaneous decay regimes in structured environments inspired by condensed matter models.

To elaborate further, let us point out that the model displayed in Fig. \ref{fig5}(a) was investigated in Ref. \cite{mouloudakis22-2} under a different perspective. Considering various spectral densities, they 
delved into the quantum Zeno effect and its connection
with an underlying
non-Hermitian Hamiltonian. 
In fact, by manipulating the master equation in Eq. (\ref{master}), 
it is easy to obtain an effective, non-Hermitian Hamiltonian by ignoring the jump term $\hat{L} \rho_{e,\ell} \hat{L}^\dagger$. This Hamiltonian features an exceptional point as discussed in Ref. \cite{mouloudakis22-2}, which is when two of the eigenvalues coalesce (the eigenvectors becoming parallel), above which the $PT$-symmetry of the Hamiltonian is broken.  
This dictates the onset of the quantum Zeno regime. We must recall, however, that the model displayed in Fig. \ref{fig5}(a) was treated here in a phenomenological manner, with no regard to the actual dynamics allowed by our disordered CCA. 
Nevertheless, there might be some CCA topologies, disordered or not, whose dynamics can exhibit features of an underlying non-Hermitian description. This is a topic worth investigating in the future.

These findings strengthen the CCA model as an excellent candidate for creating a tunable environment, enabling controlled investigation of fundamental phenomena in the dynamics of open quantum systems. Finally, recently, it has been shown that structured models, such as the discrete-time crystal model, have applications for efficient quantum computation \cite{PhysRevApplied.17.064044}. Our model might serve then as a test bed for these applications.

\section*{Acknowledgements}

We thank F. Ciccarello and P.A. Brandão for insightful discussions. This work was supported by FAPEAL (Alagoas State agency), CAPES, and CNPq (Federal agencies).

%

\end{document}